\documentclass[aps,pra,preprint,footinbib,superscriptaddress]{revtex4-1}
\usepackage{graphicx}
\usepackage{indentfirst}
\usepackage{braket}
\usepackage{float}

\usepackage{epsfig,amssymb, amsmath, mathtools}
\newcommand{\bp}{\boldsymbol \rho}
\newcommand{\bpt}{\boldsymbol{\tilde{\rho}}}

\newcommand{\bs}{{\bf s}}

\newcommand{\bzero}{{\mathbf 0}}
\newcommand{\scP}{{\mathcal P}}

\newcommand{\scE}{{\mathcal E}}

\newcommand{\scA}{{\mathcal A}}
\newcommand{\scB}{{\mathcal B}}

\newcommand{\wo}{\omega_0}

\newcommand{\wm}{\omega_-}
\renewcommand{\wp}{\omega_+}

\newcommand{\dw}{{\Delta\omega}}
\newcommand{\nm}{\,\text{nm}}

\newcommand{\m}{\,\text{m}}
\newcommand{\intI}[1]{\int\!{\rm d}#1\,}
\newcommand{\intII}[1]{\int\!{\rm d}^2#1\,}

\newcommand{\intw}[1]{\int\!\frac{{\rm d}#1}{2\pi}\,}

\newcommand{\avg}[1]{\langle #1 \rangle}
\newcommand{\bigavg}[1]{\left\langle #1 \right\rangle}
\newcommand{\abs}[1]{\left| #1 \right|}
\newcommand{\bracket}[1]{\left[ #1 \right]}
\newcommand{\parens}[1]{\left( #1 \right)}

\begin{document}
\title{Paraxial phasor-field physical optics}

\author{Justin Dove}
\email{dove@mit.edu}
\author{Jeffrey H. Shapiro}
\email{jhs@mit.edu}
\address{Research Laboratory of Electronics, Massachusetts Institute of Technology, Cambridge, MA 02139, USA}

\begin{abstract}
Phasor-field ($\scP$-field) imaging is a promising recent solution to the task of non-line-of-sight (NLoS) imaging, colloquially referred to as ``seeing around corners''. It consists of treating the oscillating envelope of amplitude-modulated, spatially-incoherent light as if it were itself an optical wave, akin to the oscillations of the underlying electromagnetic field. This resemblance enables traditional optical imaging strategies, e.g., lenses, to be applied to NLoS imaging tasks. To date, however, this ability has only been applied computationally. In this paper, we provide a rigorous mathematical demonstration that $\scP$-field imaging can be performed with physical optics, viz., that ordinary lenses can focus or project $\scP$ fields through intervening diffusers---despite these diffusers' broadly dispersing the light passing through them---and that they can image scenes hidden by such diffusers.  Hence NLoS imaging might be carried out via $\scP$-field physical optics without the nontrivial computational burden of prior NLoS techniques.
\end{abstract}

\maketitle

\section{Introduction}

Non-line-of-sight (NLoS) imaging is a growing field of research concerned with the task of generating accurate reconstructions of scenes whose observation can only be accomplished by means of intervening diffuse scattering events, e.g., by penetrating diffuse transmissive scattering media like ground glass or fog, or reflecting off diffuse surfaces like conventional walls. Traditional approaches to this problem, which largely depend on time-of-flight (ToF) information and computational backprojection, have been facilitated by pulsed illumination and time-resolved detection \cite{Kirmani2011, Velten2012}. More recent approaches have leveraged closed-form inversion techniques for ToF data \cite{O'Toole2018, lindell}, or the presence of occluders that obviate the need for ToF information \cite{Xu2018,Thrampoulidis2018}. In all of these schemes, the task of forming an image from collected data is  computational, often requiring significant overhead.

Phasor-field ($\scP$-field) imaging is a new approach to NLoS imaging that exploits the wave-like propagation behavior of the oscillating envelope of amplitude-modulated, spatially-incoherent light~\cite{Reza2018, Dove2019, Teichman2019}. It relies on the physical correlates of diffuse phase disruption often being large compared to the wavelength of the optical carrier field but insignificant relative to the much longer wavelength of radio-frequency (or even microwave) amplitude modulation. As a result, walls that diffusely scatter light, owing to their roughness at the optical-wavelength scale, appear smooth to the $\scP$ field, and thin transmissive diffusers appear transparent to the $\scP$ field. The upshot is that traditional wave-optical imaging techniques, e.g., lenses, can be applied to the $\scP$ field despite the presence of these optically disruptive elements. To date, however, this capability has only been applied computationally to form NLoS images from ToF data~\cite{Liu2019,Liu2020}.

Recently, Reza~\emph{et al.}~\cite{Reza2019} reported experiments verifying the $\scP$ field's physical wave-like properties. In one of them, they showed that a diffuse, concave reflector focuses the $\scP$ field, even though it scatters the optical carrier.  In another, they showed that an ordinary Fresnel lens focuses the optical carrier but not the $\scP$ field.  Reza~\emph{et al.}'s demonstrations led us to wonder how $\scP$-field propagation might be controlled by an ordinary lens---as opposed to the engineered diffuse reflector they used---and, in particular, is there a $\scP$-field physical-optics configuration that can enable seeing around corners in real time without the nontrivial computational burden of prior NLoS techniques. In this paper, we answer those questions through a rigorous mathematical demonstration that ordinary lenses can ndeed focus or project the $\scP$ field through intervening diffusers and that they can image a scene hidden by such diffusers.

\section{$\scP$-field setup for computational imaging}
We begin by reviewing our previously developed framework for computational $\scP$-field imaging~\cite{Dove2019}. We assume paraxial~\cite{footnoteA}, scalar-wave optics wherein an optical carrier at frequency $\wo$ is modulated by a baseband complex field envelope $E_z(\bp_z,t)$ of bandwidth $\dw \ll \wo$ to produce an optical field $U_z(\bp_z,t) = {\rm Re}[E_z(\bp_z,t) e^{-i\wo t}]$, where $\bp_z$ is the two-dimensional transverse spatial coordinate in the plane denoted by $z$. The amplitude modulation is characterized by the short-time-average (STA) irradiance, given by $I_z(\bp_z,t) = |E_z(\bp_z,t)|^2$, which can be measured by direct photodetection assuming detectors with sufficient bandwidth. The $\scP$ field is defined to be the temporal Fourier transform of the STA irradiance, averaged over any diffusers present in the scenario, whose surface fluctuations are treated statistically:
\begin{align}
	\scP_z(\bp_z,\wm) &\equiv\intI{t} \avg{I_z(\bp_z,t)} e^{i\wm t} \\ 
	&= \intw{\wp} \avg{\scE_z(\bp_z,\wp+\wm/2) \scE^*_z(\bp_z,\wp-\wm/2)},
\end{align}
where $\scE_z(\bp_z,\omega) \equiv \intI{t} E_z(\bp_z,t) e^{i\omega t}$ is the frequency-domain complex field envelope and the equality follows from $I_z(\bp_z,t) = |E_z(\bp_z,t)|^2$ plus the convolution-multiplication theorem.  In our theory the input field's complex envelope, $E_0(\bp_0,t)$, can be arbitrary, thus allowing for all manner of spatiotemporal structured illumination.  The $\scP$-field imaging that has been done to date has used fs-duration laser pulses~\cite{Liu2019,Liu2020}, but better performance (at the same pulse energy) in the physical-optics scenarios considered herein is provided by a collimated Gaussian beam with frequency-$\Omega$ double-sideband suppressed-carrier modulation, viz., 
\begin{equation}
E_0(\bp_0,t) = \sqrt{I_0}\,e^{-4|\bp_0|^2/d_0^2}e^{-t^2/T_0^2}\cos(\Omega t),\mbox{ with $\Omega T_0 \gg 1$,}
\end{equation}
because it maximizes $|\scP_0(\bp_0,2\Omega)|$ relative to the shot-noise producing background $\scP_0(\bp_0,0)$.

\begin{figure}[hbt]
\centering
\includegraphics[width=4.5in]{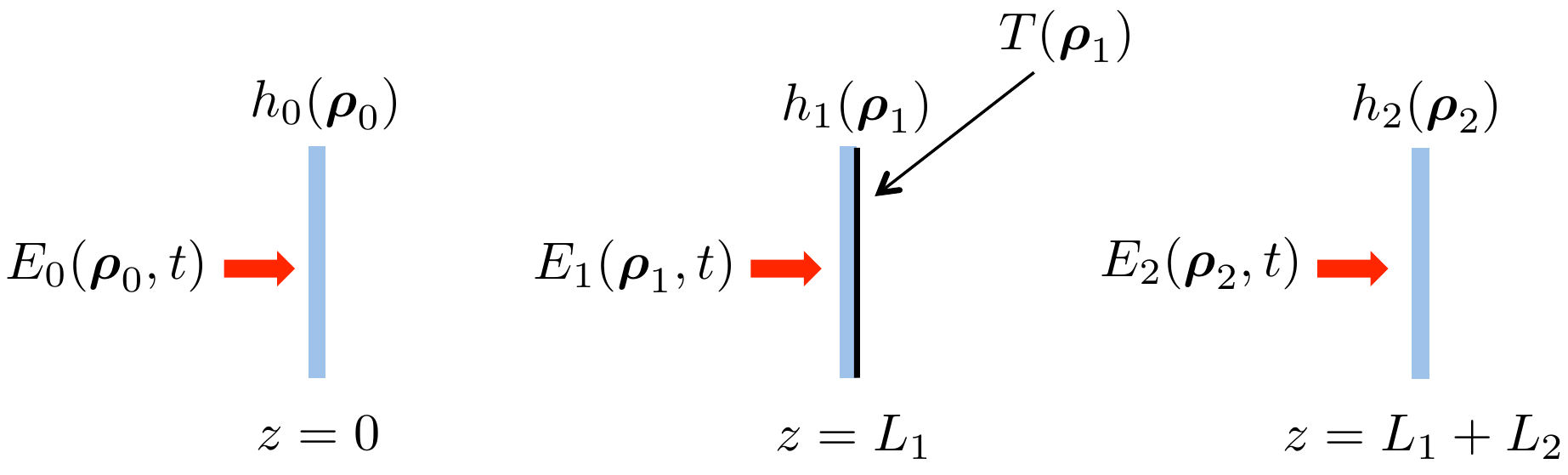}
\caption{Unfolded geometry for three-bounce NLoS active imaging. The blue rectangles represent thin transmissive diffusers, which serve as analogs of diffuse reflections, and the black line represents a thin transmissivity mask whose intensity transmission pattern, $T(\bp_1)$, represents the albedo pattern of a diffuse target in the hidden scene. Here, and elsewhere, we simplify subscripts involving $z$ where the meaning is clear from context. \label{Pfield_unfolded}}
\end{figure}

We assume a transmissive geometry, shown in Fig.~\ref{Pfield_unfolded}, that is a proxy for three-bounce reflective NLoS imaging. The frequency-domain complex envelope, $\scE'_z(\bp_z,\omega)$, that emerges from the $z$-plane diffuser when that diffuser is illuminated by $\scE_z(\bp_z,\omega)$ is given by
\begin{align}
	\scE'_z(\bp_z,\omega) = \scE_z(\bp_z,\omega) \exp\bracket{i(\wo+\omega)h_z(\bp_z)/c} \approx \scE_z(\bp_z,\omega) \exp\bracket{i \wo h_z(\bp_z)/c},
\end{align}
where the $\{h_z(\bp_z)\}$ are the diffusers' thickness profiles, which we take to be a collection of independent, identically-distributed, zero-mean Gaussian random processes with standard deviation satisfying $2\pi c/ \wo \ll \sigma_h \ll 2\pi c / \dw$, where $c$ is light speed, and correlation length obeying $\rho_h\sim 2\pi c/\wo$. Propagation of $\scE_1'(\bp_1,\omega)$ through the transmissivity mask at $z=L_1$, which is the analog of the albedo pattern of a diffuse planar target, results in
\begin{align}
	\scE''_1(\bp_1,\omega) = \scE'_1(\bp_1,\omega) \sqrt{T(\bp_1)},
\end{align}
being the complex field envelope emerging from that mask.
Free-space propagation of the complex field envelope is governed by Fresnel diffraction, i.e.,
\begin{align}
	\scE_1(\bp_1,\omega) = \frac{\wo+\omega}{2\pi i c L_1} \intII{\bp_0} \scE'_0(\bp_0,\omega) \exp\!\bracket{i(\wo+\omega)(L_1/c+|\bp_1-\bp_0|^2/2cL_1)},
\end{align}
and
\begin{align}
	\scE_2(\bp_2,\omega) = \frac{\wo+\omega}{2\pi i c L_2} \intII{\bp_1} \scE''_1(\bp_1,\omega) \exp\!\bracket{i(\wo+\omega)(L_2/c+|\bp_2-\bp_1|^2/2cL_2)},
\end{align}
where $\wo+\omega\approx\wo$ may be used in the leading factors, but not in the phase terms. The key result from our earlier work is that the $\scP$ field obeys a modified form of Fresnel diffraction when propagating away from a pure diffuser. In particular, we have that 
\begin{align}
	\scP_1(\bp_1,\wm) = \frac{1}{L_1^2} \intII{\bp_0} \scP_0(\bp_0,\wm)\exp\!\bracket{i\wm (L_1/c+|\bp_1-\bp_0|^2/2cL_1)},
	\label{P0-to-P1}
\end{align}
and
\begin{align}
	\scP_2(\bp_2,\wm) = \frac{1}{L_2^2} \intII{\bp_1} \scP_1(\bp_1,\wm) T(\bp_1) \exp\!\bracket{i\wm (L_2/c+|\bp_2-\bp_1|^2/2cL_2)},
	\label{P1-to-P2}
\end{align}
where the latter demonstrates the effect of the transmissivity mask. These results are facilitated by approximating the diffuser correlation function in integrals involving the $\scP$ field as
\begin{align}
	\bigavg{\exp\!\bracket{i \wo \parens{h_z(\bp_z) - h_z(\bp_z')}/c}} \approx \lambda_0^2\delta(\bp_z-\bp_z'),
\end{align}
where $\lambda_0=2\pi c /\wo$. In what follows, all of these results will be freely used.

Equations~(\ref{P0-to-P1}) and (\ref{P1-to-P2}) immediately imply a backprojection procedure to image the hidden albedo pattern, $T(\bp_1)$.  In particular, because
\begin{equation}
\scP_0(\bp_0,\wm) = \intw{\wp} \scE_0(\bp_0,\wp+\wm/2) \scE^*_0(\bp_0,\wp-\wm/2),
\end{equation}
is known from our choice of the illuminating field, $\scP_1(\bp_1,\wm)$ can be computed from Eq.~(\ref{P0-to-P1}).  Also, we can use conventional optics to image $I_2(\bp_2,t)$, and employ image-plane speckle averaging to obtain $\avg{I_2(\bp_2,t)}$~\cite{speckle,thesis,footnoteB}.  Then, after computing $\scP_2(\bp_2,\wm)$ from $\avg{I_2(\bp_2,t)}$, we can use backprojection to obtain $\scP_1(\bp_1)T(\bp_1)$.  

The central purpose of this paper is to show that the foregoing computational approach has a $\scP$-field physical-optics replacement.  First, in Sec.~\ref{sec:focusing}, we prove that a convex lens illuminated by a plane-wave complex field envelope can cast a focused $\scP$ field onto a plane that is hidden by an intervening diffuser which broadly scatters the optical carrier.  Next, in Sec.~\ref{sec:lens-primitive}, we establish a useful primitive for $\scP$-field propagation from an initial diffuser through a convex lens and then to an output plane.  The utility of that primitive is demonstrated in Sec.~\ref{sec:projection}, where we use it to show that a convex lens can project an arbitrary $\scP$ field from an initial diffuser to an output plane despite there being another diffuser located between the lens and the output plane.  The primitive's utility is also seen in Sec.~\ref{sec:hiddenplane-imaging}, where we employ it to show that a convex lens enables a diffuser's albedo pattern to be imaged onto a detector plane despite that diffuser's being hidden by another diffuser.

\section{Plane-wave $\scP$-field focusing} \label{sec:focusing}
\begin{figure}[hbt]
	\centering
		\includegraphics[width=4.5in]{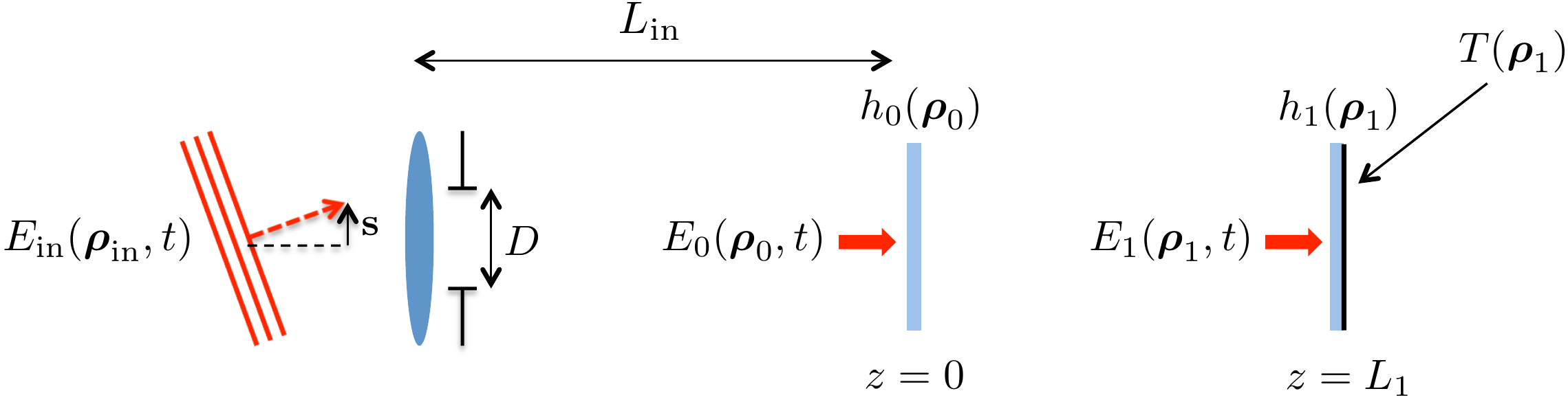}
			\caption{Geometry for focusing a plane wave through an intervening diffuser onto a target plane containing a diffuser with an albedo pattern $T(\bp_1)$. The dashed red arrow is a unit vector representing the plane wave's propagation direction, with $\bs$ being its transverse component. The lens has a Gaussian field-transmission pupil $e^{-|\bp_{\rm in}|^2/2D^2}$.\label{fig:plane-focusing}}
\end{figure}

The focusing capability of a convex lens is the natural starting point for that optical element's use in conventional physical optics.  Thus we begin our development of $\scP$-field physical optics by showing how such a lens can focus the $\scP$ field.  Consider the configuration shown in Fig.~\ref{fig:plane-focusing}, in which an infinite plane wave with complex field envelope 
\begin{equation}
E_{\rm in}(\bp_{\rm in},t) = \intw\omega \scE_{\rm in}(\omega) e^{i(\wo+\omega)\bp_{\rm in}\cdot \bs/c}e^{-i\omega t}
\end{equation}
illuminates a focal-length $f = L_{\rm in} + L_1$ lens with a Gaussian field-transmission pupil $e^{-|\bp_{\rm in}|^2/2D^2}$~\cite{footnote1}.  This field is propagating along a unit vector with transverse component $\bs$, the lens is set back a distance $L_{\rm in}$ from the first diffuser, and our goal is to focus the $\scP$ field onto the hidden target in the $z=L_1$ plane. 

The temporal-frequency-domain complex field envelope at the first diffuser is given by
\begin{equation}
	\scE_0(\bp_0,\omega) = \frac{e^{i \parens{\omega_0+\omega} L_\text{in}/c}}{i\lambda_0 L_\text{in}} \scE_\text{in}(\omega) \int\! {\rm d}^2 \bp_\text{in}\, e^{-|\bp_\text{in}|^2 / 2D^2 + i\parens{\omega_0+\omega}\parens{\bp_\text{in}\cdot\bs + |\bp_0-\bp_\text{in}|^2 / 2 L_\text{in} -|\bp_\text{in}|^2 / 2 f }/c}.
	\label{FocusStart}
\end{equation}
Performing the integral in Eq.~(\ref{FocusStart}), we get
\begin{equation}
\scE_0(\bp_0,\omega) = \frac{2\pi e^{i \parens{\omega_0+\omega} \parens{L_\text{in}+ |\bp_0|^2/2 L_\text{in} }/c}}{i\lambda_0 L_\text{in}\scA(\omega)} \scE_\text{in}(\omega) e^{-(\omega_0+\omega)^2|\bp_0-\bs L_\text{in}|^2/2c^2 L_\text{in}^2 \scA(\omega)},
\end{equation}
where
\begin{align}
	\scA(\omega) \equiv \frac{1}{D^2}-i\frac{f-L_\text{in}}{f L_\text{in}}\frac{\omega_0+\omega}{c}.
\end{align}
To simplify this result, we impose the reasonable assumption that $|{\rm Re}[\scA(\omega)]| \ll |{\rm Im}[\scA(\omega)]|$~\cite{footnoteC}. For $f=2\,\text{m}$, $L_\text{in}=1\,\text{m}$, and $\lambda_0=532\,\text{nm}$ this condition becomes $D \gg 0.4\,\text{mm}$. Using this assumption we find that
\begin{align}
	&\scE_0(\bp_0,\omega) \nonumber\\&= \frac{2\pi \scE_\text{in}(\omega)}{i\lambda_0 L_\text{in}\scA(\omega)}  e^{-|\bp_0-\bs L_\text{in}|^2 f^2/2D^2 (f-L_\text{in})^2 + i \parens{\omega_0+\omega} [L_\text{in}+|\bp_0|^2/2 L_\text{in}-|\bp_0-\bs L_\text{in}|^2 f / 2 L_\text{in} (f-L_\text{in})]/c},
\end{align}
which yields
\begin{align}
	&\scP_0(\bp_0,\omega_-) \nonumber\\&= \frac{f^2\scP_\text{in}(\omega_-)}{(f-L_\text{in})^2}  e^{-|\bp_0-\bs L_{\rm in}|^2 f^2/D^2(f-L_\text{in})^2+i \omega_- [L_\text{in}+|\bp_0|^2/2 (L_\text{in} - f) - |\bs|^2f L_\text{in}/2(f-L_\text{in}) + f\bp_0\cdot\bs/(f-L_\text{in})]/c}.
\end{align}
Using the $\scP$ field's Fresnel-diffraction formula now gives us
\begin{align}
	\lefteqn{\scP_1(\bp_1,\omega_-) = \frac{e^{i\omega_-L_1/c}}{L_1^2} \int\! {\rm d}^2 \bp_0\, \scP_0(\bp_0,\omega_-) e^{i\omega_-|\bp_1-\bp_0|^2/2 c L_1} }\\
	& = \frac{f^2\scP_\text{in}(\omega_-)}{L_1^2 (f-L_\text{in})^2}  e^{i\omega_-[L_\text{in}+L_1+|\bp_1|^2/2L_1-|\bs|^2f L_\text{in}/2(f-L_\text{in})]/c} \nonumber\\ 
	&\times \int\! {\rm d}^2 \bp_0\, e^{-|\bp_0-\bs L_{\rm in}|^2 f^2/D^2(f-L_\text{in})^2-i\omega_-[\bp_1/L_1-f\bs  /(f-L_\text{in})]\cdot\bp_0/c}e^{i \omega_- |\bp_0|^2\left(1/(L_\text{in} - f) + 1/L_1 \right)/2c}.
\end{align}
Because we have chosen $f = L_\text{in} + L_1$, the final exponential term disappears and other terms simplify. The integral that remains evaluates to 
\begin{equation}
	\scP_1(\bp_1,\omega_-) = \pi\frac{D^2}{L_1^2} e^{i\omega_- f/c} e^{i\omega_- (|\bp_1 - \bs L_{\rm in}|^2 + |\bs|^2L_{\rm in}L_1)/2cL_1} \scP_\text{in}(\omega_-) e^{-(\omega_- D/2cf)^2|\bp_1-f\bs|^2}. \label{eq:focus}
\end{equation}
If we define $\lambda_-\equiv 2\pi c/\wm$, we see that $(\wm D / 2cf)^{-1} = \lambda_- f /\pi D$. The implication of this result is that the incident plane wave creates a $\scP$-field illumination that is focused by the lens onto a diffraction-limited (at the modulation wavelength) region in the target plane whose center is offset from the origin in accord with the input illumination's angle of arrival at the lens.  

From the perspective of $\scP$-field imaging, focusing enables us to raster scan the target as if the initial diffuser were not there.  That said, although the lens focuses the $\scP$ field, it does \emph{not} focus the optical power, which is still spread out by the diffuser, as demonstrated by Reza~\emph{et al.}~\cite{Reza2019} for their diffuse, concave reflector. Because the $\scP$ field is ultimately supported by the optical field, its peak strength, like the STA irradiance's, is still subject to inverse-square law falloff, even in the presence of the lens. That Eq.~(\ref{eq:focus}) suggests otherwise, i.e., that increasing $D$ can offset the inverse-square law attenuation, is because we have assumed infinite-plane-wave illumination for which the power passing through the lens is proportional to $D^2$. Correcting for this scaling, it is clear that the peak $\scP$ field has an inverse-square law falloff relative to the input power, regardless of the pupil diameter, i.e., regardless of how tightly the $\scP$ field is confined in the target plane.

\section{Lens primitive for $\scP$-field projection and hidden-plane imaging}
\label{sec:lens-primitive}

\begin{figure}[hbt]
	\centering
	\includegraphics[width=5.5in]{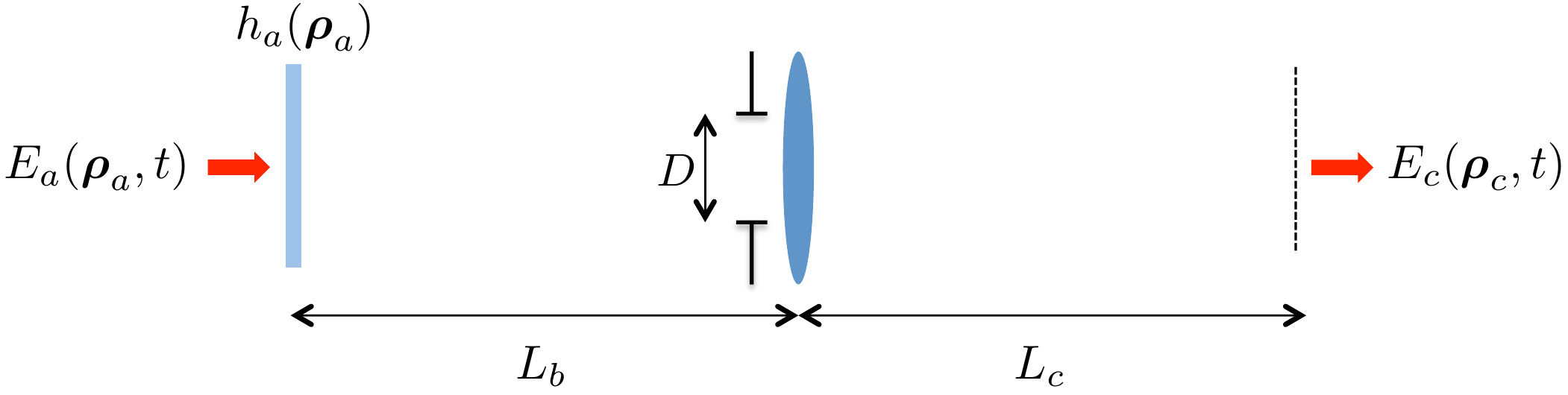}
	\caption{Geometry for post-diffuser $\scP$-field Fresnel propagation with an intervening lens.  An optical field with complex envelope $E_a(\bp_a,t)$ illuminates a diffuser $h_a(\bp_a)$ located a distance $L_b$ in front of a convex lens with focal length $f$ and a Gaussian field-transmission pupil $e^{-|\bp_b|^2/2D^2}$.  This section's objective is to relate the $\scP$ field in the plane $L_c\,$m behind the lens to the $\scP$ field entering the diffuser.} \label{fig:lens-primitive}
\end{figure}

In this section we derive a $\scP$-field propagation primitive for post-diffuser $\scP$-field Fresnel diffraction through an intervening focal-length-$f$ thin lens that has a Gaussian field-transmission pupil $e^{-|\bp_b|^2/2D^2}$, as depicted in Fig.~\ref{fig:lens-primitive}. This primitive will prove useful for both the $\scP$-field projection and $\scP$-field imaging cases that follow. We have that
\begin{eqnarray}
	\scE_c(\bp_c,\omega) &=& \frac{e^{i(\omega_0+\omega)(L_b+L_c)/c}}{-\lambda_0^2 L_b L_c} \int\! {\rm d}^2 \bp_b\, e^{-|\bp_b|^2/2D^2+i(\omega_0+\omega)\parens{|\bp_c-\bp_b|^2/L_c-|\bp_b|^2/f}/2c} \nonumber\\  &&\times\int\! {\rm d}^2 \bp_a\, \scE_a(\bp_a,\omega) e^{i(\omega_0+\omega)\parens{|\bp_b-\bp_a|^2/2L_b+h_a(\bp_a)}/c} \\
	&=& \frac{e^{i(\omega_0+\omega)\parens{L_b+L_c+|\bp_c|^2/2L_c}/c}}{-\lambda_0^2 L_b L_c} \int\! {\rm d}^2 \bp_a\, \scE_a(\bp_a,\omega) e^{i(\omega_0+\omega)\parens{|\bp_a|^2/2L_b+h_a(\bp_a)}/c}  \nonumber\\ &&\times\frac{2\pi}{\scB(\omega)} e^{-(\omega_0+\omega)^2 \left|\bp_a/L_b + \bp_c/L_c\right|^2/2c^2 \scB(\omega)},
\end{eqnarray}
where
\begin{align}
	\scB(\omega)\equiv\frac{1}{D^2}-i\frac{1}{\Pi(f,L_b,L_c)}\frac{\omega_0+\omega}{c},
\end{align}
and
\begin{align}
	\Pi(f,L_b,L_c) \equiv \frac{1}{1/L_b+1/L_c-1/f} = \frac{f L_b L_c}{f(L_b + L_c) - L_b L_c}.
\end{align}
Similar to what was done before for plane-wave focusing, we assume $|{\rm Re}[\scB(\omega)]| \ll |{\rm Im}[\scB(\omega)]|$~\cite{footnoteD} which is satisfied in both the projection and imaging scenarios to follow for parameter values similar to those chosen for the focusing case. With this assumption we have that
\begin{align}
	\scE_c(\bp_c,\omega) =& \frac{e^{i(\omega_0+\omega)\parens{L_b+L_c+|\bp_c|^2/2L_c}/c}}{-\lambda_0^2 L_b L_c} \int\! {\rm d}^2 \bp_a\, \scE_a(\bp_a,\omega) e^{i(\omega_0+\omega)\parens{|\bp_a|^2/2L_b+h_a(\bp_a)}/c}  \nonumber\\ &\times\frac{2\pi}{\scB(\omega)} e^{-\left|\bp_a/L_b + \bp_c/L_c \right|^2\scB^*(\omega)\Pi^2(f,L_b,L_c)/2},
\end{align}
from which it follows that
\begin{align}
	\lefteqn{\scP_c(\bp_c,\omega_-) = 
	 \left(\frac{\Pi(f,L_b,L_c)}{L_b L_c}\right)^2 e^{i\omega_-\parens{L_b+L_c+|\bp_c|^2/2L_c}/c} }\nonumber\\ &\times \int\! {\rm d}^2 \bp_a\, \scP_a(\bp_a,\omega_-) e^{-\left|\bp_a/L_b + \bp_c/L_c\right|^2\Pi^2(f,L_b,L_c)/D^2}e^{i\omega_-\parens{|\bp_a|^2/L_b- \Pi(f,L_b,L_c)\abs{\bp_a / L_b +\bp_c/ L_c}^2 }/2 c}.
	\label{eq:lens-prim}
\end{align}

\section{$\scP$-field projection}
\label{sec:projection}
\begin{figure}[hbt]
	\centering
	\includegraphics[width=5.5in]{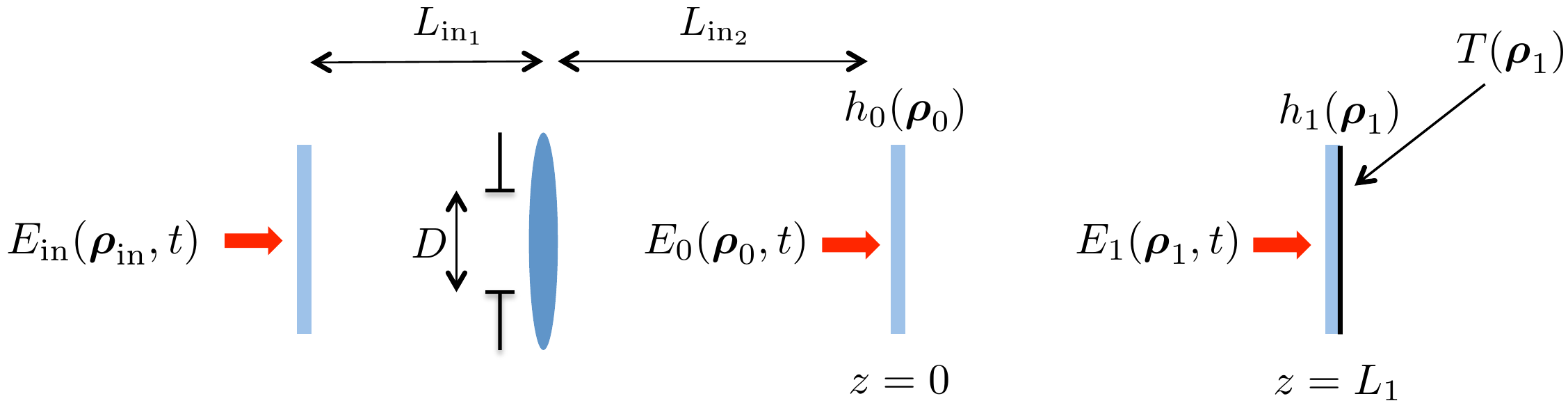}
	\caption{Geometry for projecting an arbitrary $\scP$-field pattern from an input diffuser~\cite{footnoteE} to a hidden target plane.  An optical field with complex envelope $E_{\rm in}(\bp_{\rm in},t)$ illuminates an instance of the Fig.~\ref{fig:lens-primitive} geometry in which $E_{\rm in}(\bp_{\rm in},t)$ takes the place of $E_a(\bp_a,t)$, and $L_{{\rm in}_1}$, $L_{{\rm in}_2}$ take the places of $L_b$, $L_c$, respectively. The lens, whose focal length satisfies $1/f = 1/L_{{\rm in}_1} + 1/(L_{{\rm in}_2}+L_1)$, then projects the $\scP$ field emerging from the initial diffuser onto the $z=L_1$ plane with magnification/minification factor $M = L_{\rm prj}/L_{{\rm in}_1}$, where $L_{\rm prj} \equiv L_{{\rm in}_2} + L_1$.} \label{fig:projecting}
\end{figure}

With the Eq.~(\ref{eq:lens-prim}) primitive in hand, we now turn our attention to the task of projecting an arbitrary $\scP$-field pattern from an input diffuser to a hidden target plane. This arrangement will allow structured-illumination paradigms for line-of-sight imaging with coherent light to be translated into structured-illumination NLoS imaging with $\scP$ fields.

Consider an augmentation of the Fig.~\ref{Pfield_unfolded} scenario in which that figure's initial diffuser is preceded by an instance of the lens primitive depicted in Fig.~\ref{fig:lens-primitive}, as shown in Fig.~\ref{fig:projecting}. Modifying the Fig.~\ref{fig:lens-primitive} scenario's placeholder notation, we will label the transverse coordinate of Fig.~\ref{fig:projecting}'s input plane as $\bp_\text{in}$, its first distance as $L_{\text{in}_1}$, and its second distance as $L_{\text{in}_2}$. The lens primitive's output-plane transverse coordinate remains as $\bp_0$, leading into the same notation as Fig.~\ref{Pfield_unfolded} for the rest of that figure's geometry. We take the lens to be configured to project the input $\scP$ field onto the $z=L_1$ plane by choosing its focal length to obey $1/f = 1/L_{\text{in}_1} + 1/(L_{\text{in}_2}+L_1)$. For this configuration we have $\Pi(f,L_{\text{in}_1},L_{\text{in}_2}) = L_{\text{in}_2}(L_1+L_{\text{in}_2})/L_1$, and so the lens primitive gives us
\begin{align}
	\scP_0(\bp_0,\omega_-) =& \left(\frac{L_1+L_{\text{in}_2}}{L_1 L_{\text{in}_1}}\right)^2 e^{i\omega_-\parens{L_{\text{in}_1}+L_{\text{in}_2}-|\bp_0|^2/2L_1}/c} \nonumber\\ &\times\int\! {\rm d}^2 \bp_\text{in}\, \scP_\text{in}(\bp_\text{in},\omega_-) e^{-\left|\bp_\text{in}/L_{\text{in}_1} + \bp_0/L_{\text{in}_2}\right|^2\left(L_{\text{in}_2}(L_1+L_{\text{in}_2})/D L_1\right)^2} \nonumber\\ &\times e^{i\omega_-\bracket{|\bp_\text{in}|^2\left(1-L_{\text{in}_2}(L_1+L_{\text{in}_2}) / L_1 L_{\text{in}_1}\right)-2\bp_\text{in}\cdot\bp_0(L_1+L_{\text{in}_2}) / L_1 } / 2 c L_{\text{in}_1}},
\end{align}
which, after Fresnel propagation, leads to
\begin{align}
	\scP_1(\bp_1,\omega_-) &= \left(\frac{L_1+L_{\text{in}_2}}{L_1^2 L_{\text{in}_1}}\right)^2 e^{i\omega_-\parens{L_{\text{in}_1}+L_{\text{in}_2}+L_1+|\bp_1|^2/2L_1}/c} \nonumber\\ & \hspace*{.2in}\times \int\! {\rm d}^2 \bp_\text{in}\, \scP_\text{in}(\bp_\text{in},\omega_-) e^{i\omega_-|\bp_\text{in}|^2 \left(1-L_{\text{in}_2}(L_1+L_{\text{in}_2})/L_1 L_{\text{in}_1}\right)/2 c L_{\text{in}_1}} \nonumber\\ &\hspace*{.2in}\times\int\! {\rm d}^2 \bp_0\, e^{-\left|\bp_\text{in} / L_{\text{in}_1} + \bp_0 / L_{\text{in}_2}\right|^2\left(L_{\text{in}_2}(L_1+L_{\text{in}_2}) / D L_1\right)^2-i\omega_-\bp_0\cdot\left(\bp_1+\bp_\text{in}(L_1+L_{\text{in}_2}) / L_{\text{in}_1} \right) / c L_1} \\
	&= \pi \left(\frac{D}{L_1 L_{\text{in}_1}}\right)^2 e^{i\omega_-\parens{L_{\text{in}_1}+L_{\text{prj}}+|\bp_1|^2/2L_{\text{prj}}}/c} \nonumber\\ & \hspace*{.2in}\times\int\! {\rm d}^2 \bp_\text{in}\, \scP_\text{in}(\bp_\text{in},\omega_-) e^{i\omega_-|\bp_\text{in}|^2/2cL_{\text{in}_1}} e^{-\left|\bp_1+M \bp_\text{in}\right|^2[(\omega_- D / 2 c L_{\text{prj}})^2-i\wm L_{\text{in}_2} / 2 c L_1 L_{\text{prj}}]}  ,
	\label{eq:proj-mid}
\end{align}
where $L_\text{prj}\equiv L_{\text{in}_2}+L_1$ and $M\equiv L_\text{prj}/L_{\text{in}_1}$ is the magnification/minification factor. Now we assume a more stringent condition for $D$ than what was needed to ensure $|{\rm Re}[\mathcal{B}(\omega)]| \ll  |{\rm Im}[\mathcal{B}(\omega)]|$, viz., $D \gg \sqrt{c L_{\text{in}_2}L_{\text{prj}}/\wm L_1}$. For meter-scale distances and 10-GHz-scale modulation this reduces to approximately $D \gg 10\,\text{cm}$ which, although likely difficult to meet in practice, is at least imaginable, perhaps by using a large concave mirror to function as the lens. Each order-of-magnitude increase of the modulation frequency reduces the requirement on $D$ by half an order of magnitude, so THz-scale modulation---as implemented by Willomitzer~\emph{et al.}'s synthetic-wavelength holography~\cite{smu}, which we discuss in Appendix~\ref{app:hologram}---would reduce this condition to a more reasonable $D \gg 1\,\text{cm}$. If we can achieve this condition, then the final phase term in Eq.~(\ref{eq:proj-mid}) can be ignored and we get
\begin{align}
	\scP_1(\bp_1,\omega_-) =& \pi \left(\frac{D}{L_1 L_{\text{in}_1}}\right)^2 e^{i\omega_-(L_{\text{in}_1}+L_{\text{prj}})/c} e^{i\omega_- |\bp_1|^2/2cL_{\text{prj}}} \nonumber\\&\times\int\! {\rm d}^2 \bp_\text{in}\, \scP_\text{in}(\bp_\text{in},\omega_-) e^{i\omega_-|\bp_\text{in}|^2/2cL_{\text{in}_1}} e^{-(\omega_- D/2cL_\text{prj})^2\left|\bp_1+M \bp_\text{in}\right|^2}.
\end{align}
Ignoring inessential phase and scaling terms, this is a projected copy of the input $\scP$ field,
\begin{equation}
\mathcal{P}_\text{in}(\bp_\text{in},\omega_-) = \int\!\frac{{\rm d}\omega_+}{2\pi}\,
\mathcal{E}_\text{in}(\bp_\text{in},\omega_+ +\omega_-/2)\mathcal{E}^*_\text{in}(\bp_\text{in},\omega_+ -\omega_-/2),
\end{equation}
subject to image inversion and magnification/minification, with resolution diffraction limited at the modulation wavelength. This result enables structured illumination~\cite{struct} and dual photography~\cite{dual} techniques to be applied to NLoS $\scP$-field imaging.

\section{$\scP$-field imaging of a hidden plane}
\label{sec:hiddenplane-imaging}

\begin{figure}[hbt]
	\centering
	\includegraphics[width=4.5in]{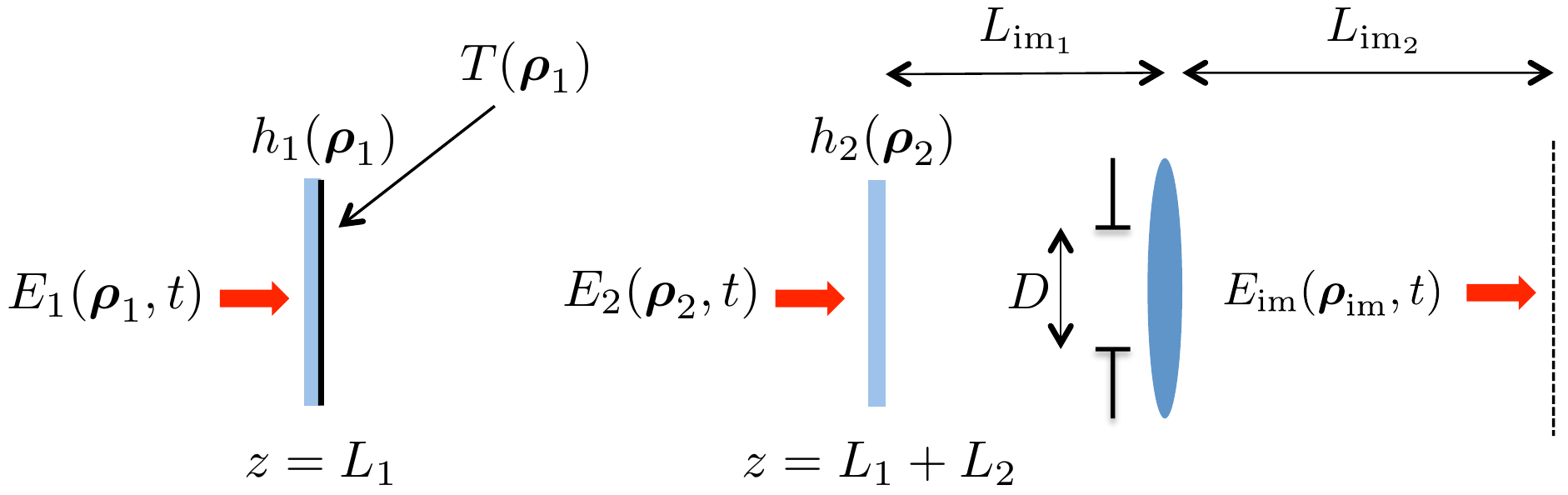}
	\caption{Geometry for physical-optics $\scP$-field imaging of a hidden target plane.  The complex field envelope $E_1(\bp_1,t)$ is the hidden-plane illumination in Fig.~\ref{Pfield_unfolded} that results when $E_0(\bp_0,t)$ illuminates a diffuser in the $z=0$ plane. The rest of the setup shown here is an instance of the lens primitive from Fig.~\ref{fig:lens-primitive} in which $E_2(\bp_2,t)$ takes the place of $E_a(\bp_a,t)$ and $L_{{\rm in}_1}$, $L_{{\rm in}_2}$ take the places of $L_b$, $L_c$, respectively.  The lens, whose focal length satisfies $1/f = 1/(L_2+L_{{\rm im}_1}) + 1/L_{{\rm im}_2}$, then casts an image of $\scP_1(\bp_1,t)T(\bp_1)$ onto the $z= L_1+L_2+L_{{\rm im}_1}+L_{{\rm im}_2}$ plane with minification/magnification factor $M = L_{\rm out}/L_{{\rm im}_2}$, where $L_{\rm out} \equiv L_2 + L_{{\rm im}_1}.$}\label{fig:imaging}
\end{figure}

It turns out that the same kind of result just exhibited for $\scP$-field projection can be obtained for physical-optics $\scP$-field imaging of a hidden target plane.  To do so, we place our lens primitive from Fig.~\ref{fig:lens-primitive} behind Fig.~\ref{Pfield_unfolded}'s final diffuser so that the transverse coordinate of the first plane of the primitive is $\bp_2$, as depicted in Fig.~\ref{fig:imaging}. We denote the lens primitive's two distances as $L_{\text{im}_1}$ and $L_{\text{im}_2}$, and we denote the transverse coordinate of the primitive's final plane as $\bp_\text{im}$ with the intention in mind that a multipixel detector array capable of measuring the $\scP$ field will be present there. The hidden plane is imaged onto this hypothetical detector by taking the focal length of the lens to obey $1/f=1/(L_2+L_{\text{im}_1})+1/L_{\text{im}_2}$. For this configuration we have $\Pi(f,L_{\text{im}_1},L_{\text{im}_2}) = L_{\text{im}_1}(L_2+L_{\text{im}_1})/L_2$, and so we get
\begin{align}
	\scP_\text{im}(\bp_\text{im},\omega_-)=& \left(\frac{L_2+L_{\text{im}_1}}{L_2 L_{\text{im}_2}}\right)^2 e^{i\omega_-\parens{L_{\text{im}_1}+L_{\text{im}_2}+|\bp_\text{im}|^2\left(1-L_{\text{im}_1}(L_2+L_{\text{im}_1})/L_2 L_{\text{im}_2}\right)/2L_{\text{im}_2}}/c}  \nonumber\\ &\times \int\! {\rm d}^2 \bp_2\, \scP_2(\bp_2,\omega_-) e^{-\left|\bp_2 / L_{\text{im}_1} + \bp_\text{im} / L_{\text{im}_2}\right|^2\left(L_{\text{im}_1}(L_2+L_{\text{im}_1}) / D L_2 \right)^2 } \nonumber\\ &\times e^{- i\omega_-[|\bp_2|^2 / 2 +  \bp_\text{im}\cdot\bp_2(L_2+L_{\text{im}_1}) / L_{\text{im}_2}] / c L_2} \\
	=& \frac{M^2}{L_2^4} e^{i\omega_-\parens{L_{\text{out}}+L_{\text{im}_2}+|\bp_\text{im}|^2\left(1-M L_{\text{im}_1}/L_2\right)/2L_{\text{im}_2}}/c} \int\! {\rm d}^2 \bp_1\, T(\bp_1) \scP_1(\bp_1,\omega_-)  \nonumber\\ & \times e^{i\omega_-|\bp_1|^2/2 c L_2}\int\! {\rm d}^2 \bp_2\,  e^{-\left|\bp_2/L_{\text{im}_1} + \bp_\text{im}/L_{\text{im}_2}\right|^2\left(L_{\text{im}_1}L_\text{out}/D L_2\right)^2 - i\omega_-\bp_2\cdot\left(\bp_1+M \bp_\text{im}\right)/c L_2} \\ 
	=& \pi \left(\frac{D}{L_2 L_{\text{im}_2}}\right)^2 e^{i\omega_-\parens{L_{\text{out}}+L_{\text{im}_2}+|\bp_\text{im}|^2/2L_{\text{im}_2}}/c}  \int\! {\rm d}^2 \bp_1\, T(\bp_1) \scP_1(\bp_1,\omega_-) e^{i\wm|\bp_1|^2/2cL_\text{out}}\nonumber\\ & \times e^{-\left|\bp_1 + M \bp_\text{im}\right|^2\parens{ \parens{D\omega_-/2c L_{\text{out}}}^2 - i\omega_- L_{\text{im}_1}/ 2 c L_2 L_{\text{out}}}},
\end{align}
where $L_\text{out}\equiv L_2+L_{\text{im}_1}$ and $M\equiv L_\text{out} / L_{\text{im}_2}$ is the minification/magnification factor. Similar to the projection case, we enforce the assumption $D \gg \sqrt{c L_{\text{im}_1}L_{\text{out}}/\wm L_2}$ so that we can ignore the final phase term, which leaves us with
\begin{align}
	\scP_\text{im}(\bp_\text{im},\omega_-) =& \pi \left(\frac{D}{L_2 L_{\text{im}_2}}\right)^2 e^{i\omega_-(L_{\text{out}}+L_{\text{im}_2})/c} e^{i\omega_-|\bp_\text{im}|^2/2cL_{\text{im}_2}} \nonumber\\ &\times\int\! {\rm d}^2 \bp_1\, T(\bp_1) \scP_1(\bp_1,\omega_-) e^{i\omega_- |\bp_1|^2/2cL_\text{out}} e^{-(\omega_- D/2cL_\text{out})^2\left|\bp_1 + M\bp_\text{im}\right|^2}.
	\label{eq:imaging}
\end{align}
Again ignoring inessential terms, this is an inverted and minified/magnified $\scP$-field image of the hidden target plane with spatial resolution that is diffraction limited at the modulation wavelength. This result---in our transmissive proxy for three-bounce NLoS imaging---points to the potential for direct $\scP$-field imaging of NLoS scenes with little or no computational overhead. Note that our physical-optics $\scP$-field imager uses its lens to directly image the hidden plane at $z=L_1$ in Fig.~\ref{Pfield_unfolded}. In contrast, for the same Fig.~\ref{Pfield_unfolded} geometry the computational $\scP$-field imager we considered in Ref.~\cite{Dove2019} reconstructs the hidden plane's albedo pattern from an image of the $z=L_1+L_2$ plane, analogous to what is done in three-bounce NLoS experiments that use laser illumination.  

\section{Summary and discussion}
In this paper, we have mathematically investigated the possibility for NLoS imaging tasks to be carried out in the $\scP$-field framework using physical optics in lieu of the more involved computational techniques employed to date. Using an unfolded proxy for three-bounce NLoS imaging and assuming paraxial propagation, we first demonstrated that plane-wave $\scP$-field illumination can be focused onto a small target-plane region despite the presence of an intervening diffuser.  Moreover, this $\scP$-field focusing occurs even though the optical carrier, and thus the power, is broadly scattered. Next, we developed a useful primitive for $\scP$-field propagation in scenarios involving lenses.  We then applied that primitive to $\scP$-field projection and $\scP$-field imaging of a hidden plane's albedo pattern. We found that arbitrary $\scP$-field patterns can be projected, through an intervening diffuser, onto a target plane.  Likewise, we showed that $\scP$-field imaging of a target plane's albedo pattern can be accomplished despite the presence of another intervening diffuser between that plane and the sensor. In these last two cases, the mathematical assumptions made in our analysis rely on rather large lenses. However, the qualifying lens size can be reduced by increasing the $\scP$-field frequency. For all three cases---$\scP$-field focusing, projection, and imaging---we found that the spatial resolution was at the $\scP$-field frequency's diffraction limit, further encouraging the pursuit of higher $\scP$-field frequencies.  

Ultimately, the highest usable $\scP$-field frequency for physical-optics imaging will be set by the bandwidth of available detectors, which is not likely to exceed 100\,GHz. Despite this detector limitation, we allege that higher $\scP$-field frequencies, perhaps up to 1\,THz, can be achieved using Willomitzer~\emph{et al.}'s synthetic-wavelength holography~\cite{smu}, wherein coherently-detected outputs from sequentially illuminating an NLoS system with unmodulated light at two optical frequencies can be correlated and filtered to produce a synthetic $\scP$ field at the difference frequency, without the need for ultrafast detectors. This technique offers a computational approach to $\scP$-field generation and detection that stands in contrast to this paper's physical-optics paradigm. However, the required computation is trivial in comparison to what is typically used, at present, for extracting images from NLoS datasets. So, even for this synthetic approach, the conclusion of this paper stands:  there is real potential for NLoS imaging to be accomplished with traditional optical techniques in the $\scP$-field framework using real, physical optics, obviating the need for the nontrivial computational inversion schemes that have been applied to date. Nevertheless, much more needs to be done to bring $\scP$-field physical optics NLoS imaging to fruition.  For one thing, many NLoS geometries will violate the validity conditions for paraxial propagation, thus requiring use of Rayleigh--Sommerfeld diffraction for accurate scene reconstructions from $\scP$-field measurements.  We have already taken initial steps~\cite{thesis,Nonparaxial} in this direction, in the transmissive geometry, but additional work is needed to convert them to the three-bounce NLoS scenario. For another, our transmissive geometry addresses 2D imaging, i.e., recovering the albedo pattern present on a hidden planar diffuser, but NLoS imaging almost invariably involves reconstructing a hidden 3D scene.  Consequently, for NLoS imagers depth-of-focus and depth-of-field considerations come into play.  We expect that these issues can be addressed in our $\scP$-field framework, but that is a subject for future work.

\appendix 

\section{Synthetic-wavelength holography}
\label{app:hologram}

In this appendix, we will develop Willomitzer~\emph{et al.}'s  synthetic-wavelength holography~\cite{smu} in our $\scP$-field framework. Within the limitations of current laser and detector technology, we will show that synthetic-wavelength holography (SWH) offers a practical method to access higher $\scP$-field frequencies than are presently attainable with modulated illumination and direct detection. 

\begin{figure}[h]
	\centering
	\includegraphics[width=5.2in]{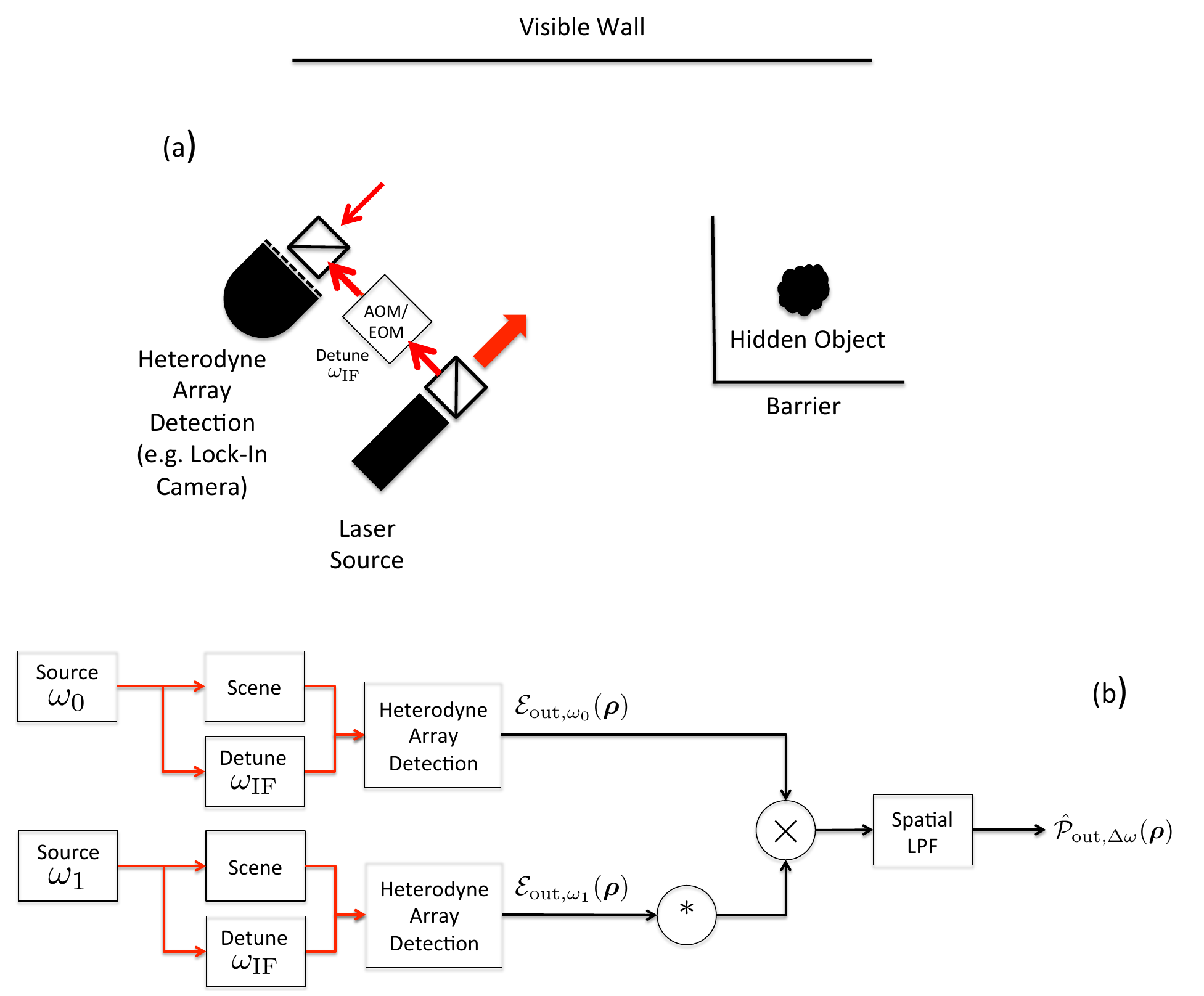}
	\caption{Synthetic wavelength holography (SWH).  (a) Setup for NLoS SWH imaging as performed by Willomitzer~\emph{et al}.~\cite{smu}. A tunable continuous-wave laser sequentially illuminates a spot on the visible wall at frequencies $\omega_0$ and $\omega_1 = \omega_0-\Delta\omega$.  Diffuse reflection  from that wall illuminates a hidden object.  Diffuse reflection from that object illuminates the visible wall and a portion of that wall---disjoint from the original illumination spot---is imaged onto a heterodyne detection array.  That array's local oscillator is obtained by acousto-optic modulation (AOM) or electro-optic modulation (EOM) that produces an intermediate-frequency shift $\omega_{\rm IF}$.  The array uses, e.g., a lock-in camera that outputs the pixel-wise quadrature components of the IF signals that are used to generate a computational image of the hidden object. (b)  Measurement and signal-processing chain for SWH imaging in our transmissive-geometry proxy for three-bounce NLoS imaging. Ideal heterodyne detection is assumed, with a continuum array whose signal-to-noise ratio is high enough that its outputs can be taken to be the complex field envelopes, $\scE_{{\rm out},\omega_0}(\bp)$ and $\scE_{{\rm out},\omega_1}(\bp)$, of the imaged fields.  Multiplying 
$\scE_{{\rm out},\omega_0}(\bp)$ by the complex conjugate of $\scE_{{\rm out},\omega_1}(\bp)$ and passing that product through a spatial low-pass filter (LPF) then yields $\hat{\scP}_{{\rm out},\Delta\omega}(\bp)$, i.e., an estimate of the desired $\scP$ field.} \label{fig:SWH}
\end{figure}

Figure~\ref{fig:SWH}(a) shows the physical setup for SWH NLoS imaging.  A tunable continuous-wave laser sequentially illuminates a spot on a visible wall at frequencies $\omega_0$ and $\omega_1 = \omega_0 - \Delta \omega$.  Diffuse reflection from the visible wall then illuminates an object in the hidden space and diffuse reflections therefrom return to a large swath of the visible wall. Light reflected from a visible-wall patch that is disjoint from the laser-illuminated spot is imaged onto a heterodyne-detection array.  This array gets its local-oscillator field from acousto-optic modulation (AOM) or electro-optic modulation (EOM) of the transmitter laser that produces an intermediate-frequency (IF) offset $\omega_{\rm IF}$.  The array uses a lock-in camera (or equivalent) that outputs the pixel-wise quadrature components of the IF signals.  These pixel-wise quadratures are subsequently processed to generate a computational image of the hidden object.

Willomitzer~\emph{et al}.\@ drew their inspiration from Gabor's Nobel-prize winning work on holography, and their analysis/synthesis approach to SWH does not rely on statistics.  Our objective, in this appendix, is to flesh out the $\scP$-field SWH measurement and signal-processing chain shown in Fig.~\ref{fig:SWH}(b) to show how our statistical framework reproduces the basic concept from their paper in our transmissive-geometry proxy for NLoS imaging. 

To begin, consider the $\scP$ field associated with single-frequency irradiance modulation at angular frequency $\dw$:
\begin{align}
	\scP(\bp,\wm) = \scP^*_{\dw}(\bp)2\pi\delta(\wm+\dw) + \scP_{0}(\bp)2\pi\delta(\wm) + \scP_{\dw}(\bp)2\pi\delta(\wm-\dw).
\end{align}
One way to generate this $\scP$ field is by coherently summing optical fields at two angular frequencies, $\wo$ and $\omega_1 \equiv \wo - \dw$, in which case the frequency-domain complex field envelope associated with such a signal is 
\begin{align}
	\scE(\bp,\omega) = \scE_{\omega_0}(\bp)2\pi\delta(\omega) + \scE_{\omega_1}(\bp)2\pi\delta(\omega+\dw).
\end{align}
The $\scP$ field associated with this complex field envelope is given by
\begin{align}
	\scP(\bp,\wm) =& \intw\wp \avg{\scE(\bp,\wp+\wm/2)\scE^*(\bp,\wp-\wm/2)} \\
	=& \avg{\scE^*_{\omega_0}(\bp)\scE_{\omega_1}(\bp)}2\pi\delta(\wm+\dw) + (\avg{\abs{\scE_{\omega_0}(\bp)}^2}+ \avg{\abs{\scE_{\omega_1}(\bp)}^2}) 2\pi\delta(\wm) \nonumber\\&+ \avg{\scE_{\omega_0}(\bp)\scE^*_{\omega_1}(\bp)}2\pi\delta(\wm-\dw).
\end{align}
This is a single-frequency $\scP$ field at frequency $\dw$. In particular, we have that
\begin{subequations}
\begin{align}
	\scP_{0}(\bp) &= \avg{\abs{\scE_{\omega_0}(\bp)}^2}+ \avg{\abs{\scE_{\omega_1}(\bp)}^2}\\
	\scP_\dw(\bp) &= \avg{\scE_{\omega_0}(\bp)\scE^*_{\omega_1}(\bp)}.
\end{align}
	\label{eq:mono-components}
\end{subequations}

For systems that are linear and time invariant with respect to the optical field---as is the case for typical NLoS scenarios---the frequency structure of the $\scP$ field and its underlying complex field envelope are unaffected by that system.  Consequently, propagating them through that system reduces to relating the input quantities to the output quantities at each frequency. It follows from Eq.~(\ref{eq:mono-components}) that the input-output relations for the complex field envelopes at $\omega_0$ and $\omega_1$ suffice to characterize the input-output behavior of the $\scP$ field. More importantly, for linear time-invariant systems, the input-output relations that take inputs  $\scE_{{\rm in},\omega_0}(\bp_{\rm in})$ and $\scE_{{\rm in},\omega_1}(\bp_{\rm in})$ and yield outputs $\scE_{{\rm out},\omega_0}(\bp_{\rm out})$ and $\scE_{{\rm out},\omega_1}(\bp_{\rm out})$ are the same as the complex field envelopes' input-output relations for illuminating the system with unmodulated light at each frequency \emph{individually}. The implication is that, provided we can accurately measure the output complex field envelopes and accomplish the required diffuser averaging, phasor-field imaging tasks can be carried out by \emph{sequentially} illuminating the system with unmodulated inputs, meaning we are not burdened by needing direct detectors that are sufficiently fast to capture the $\scP$-field modulation frequency. We will turn now to each of these concerns where, for convenience, we will use $\bp$ instead of $\bp_{\rm out}$ for the output plane's coordinate vector.

To measure the output complex field envelope $\scE_{{\rm out},\omega_n}(\bp)$, for $n=0,1$, we will use balanced heterodyne detection~\cite{heterodyne,footnoteF}. The signal field is mixed on a 50--50 beam splitter with a strong, plane-wave local oscillator that is detuned from $\omega_n$ by an intermediate frequency $\omega_{\rm IF} \ll \Delta\omega$ and has phase $\theta_{\rm LO}$. The light emerging from the beam-splitter's output arms are detected, at high spatial resolution, with detector arrays whose temporal bandwidths, including those of their post-detection electronics, exceed $\omega_{\rm IF}$. For simplicity, we will treat these arrays as performing ideal continuum photodetection, i.e., they have unlimited spatial resolution. We will also ignore the effects of noise and assume the arrays have unity quantum efficiency. In effect, we take the arrays to accurately detect the STA irradiances at the beam-splitter outputs---denoted $I_+(\bp,t)$ and $I_-(\bp,t)$---at full spatial resolution. The difference of these outputs carries the desired complex field envelope at the intermediate frequency:
\begin{align}
	I_+(\bp,t)-I_-(\bp,t) \propto {\rm Re}\!\bracket{\scE_{{\rm out},\omega_n}(\bp) e^{-i(\omega_{\rm IF} t - \theta_{\rm LO})}},
\end{align}
when the illumination frequency is $\omega_n$, for $n=0,1$. Thus the quadratures of that signal can be extracted by standard communication electronics and we obtain $\scE_{{\rm out},\omega_n}(\bp)$. 

Now we can computationally form $\scE_{\text{out},\omega_0}(\bp)\scE^*_{\text{out},\omega_1}(\bp)$, but we still need to perform diffuser averaging. It is inadvisable to use arrays whose detector elements average over many speckles, because that destroys the signal of interest, i.e., 
\begin{align}
	\bigavg{I_+(\bp,t)-I_-(\bp,t)} \propto {\rm Re}\!\bracket{\avg{\scE_{{\rm out},\omega_n}(\bp)} e^{-i(\omega_{\rm IF} t - \theta_{\rm LO})}} = 0, \mbox{ for $n=0,1$}.
\end{align}
Hence, it is critical that we use detector arrays whose spatial resolution is high enough to resolve the speckle pattern. In practice, we want no more than a few speckle cells to fall on each pixel in our detector arrays to avoid the associated signal attenuation. So, to perform diffuser averaging, we propose low-pass spatial filtering $\scE_{\text{out},\omega_0}(\bp)\scE^*_{\text{out},\omega_1}(\bp)$ with a Gaussian kernel to generate 
\begin{align}
	\hat\scP_{\text{out},\dw}(\bp) \equiv \frac{1}{2\pi R^2} \intII\bpt e^{-\abs{\bp-\bpt}^2/2R^2}\scE_{\text{out},\omega_0}(\bpt)\scE^*_{\text{out},\omega_1}(\bpt).
\end{align}
Provided that our Gaussian kernel resolves the spatial features in the $\scP$ field, viz., $R < \Delta\lambda \equiv 2\pi c / \dw$, then $\hat\scP_{\text{out},\dw}(\bp)$ is approximately unbiased, i.e., 
\begin{align}
	\avg{\hat\scP_{\text{out},\dw}(\bp)} &= \frac{1}{2\pi R^2} \intII\bpt e^{-\abs{\bp-\bpt}^2/2R^2}\avg{\scE_{\text{out},\omega_0}(\bpt)\scE^*_{\text{out},\omega_1}(\bpt)} \\
	&= \frac{1}{2\pi R^2} \intII\bpt e^{-\abs{\bp-\bpt}^2/2R^2}\scP_{\text{out},\dw}(\tilde{\bp}) \\
	&\approx \scP_{\text{out},\dw}(\bp). 
\end{align}

What remains then is to assess the variance of $\hat\scP_{\text{out},\dw}(\bp)$. To do so, we appeal to a speckle analysis developed elsewhere~\cite{thesis, speckle}. We assume the unfolded three-bounce geometry from Fig.~\ref{Pfield_unfolded}, with monochromatic (frequency $\omega_n$) illumination of power $P$ with spatial profile $\scE_0(\bp_0) = \sqrt{8P/\pi d_0^2} \exp(-4|\bp_0|^2/d_0^2)$. For simplicity, the target at the second bounce is replaced by a Gaussian field-transmissivity pupil, $\exp(-4|\bp_1|^2/d_1^2)$.  Another such pupil, $\exp(-4|\bp_2|^2/d_2^2)$, is placed at the third-bounce location to characterize the finite size of the visible wall, and the distances propagated after each bounce are all $L$. We shall also assume that the Fresnel number products, $d_0^2 d_1^2 / \lambda_0^2 L^2$ and $d_1^2 d_2^2 / \lambda_0^2 L^2$, are large enough that the third-bounce speckle is well-approximated as single-bounce speckle from the final diffuser~\cite{speckle}.  This condition leads to a third-bounce complex field envelope that is Gaussian distributed.  Gaussian moment factoring then gives us
\begin{align}
	\text{var}(\hat\scP_{\text{out},\dw}(\bp)) &= \avg{|\hat\scP_{\text{out},\dw}(\bp)|^2} - |\avg{\hat\scP_{\text{out},\dw}(\bp)}|^2 \\ &= \frac{1}{4\pi^2 R^4}\intII{\bp_\text{out}} \intII{\bpt_\text{out}} e^{-(\abs{\bp-\bp_\text{out}}^2 + \abs{\bp-\bpt_\text{out}}^2)/2R^2}\nonumber \\ &\hspace*{.2in}\avg{\scE_{\text{out},\omega_0}(\bp_\text{out})\scE^*_{\text{out},\omega_0}(\bpt_\text{out})} \avg{\scE^*_{\text{out},\omega_1}(\bp_\text{out})\scE_{\text{out},\omega_1}(\bpt_\text{out})},
\end{align}
where $\mathcal{E}_{{\rm out},\omega_n}(\bp_{\rm out})$---the spatial profile in the plane that lies a distance $L$ beyond Fig.~\ref{Pfield_unfolded}'s output plane---satisfies
\begin{align}
	\scE_{{\rm out},\omega_n}(\bp_{\rm out}) = \frac{\omega_n}{i2\pi c L} e^{i\omega_n L/c} \intII{\bp_2} \scE_{2,\omega_n}(\bp_2) e^{i\omega_n\abs{\bp_{\rm out}-\bp_2}^2 / 2 c L} e^{i \omega_n h_2(\bp_2) / c} e^{-4\abs{\bp_2}^2/d_2^2},
\end{align}
and similar relations hold for the previous bounces. Accordingly, we find that
\begin{align}
	\avg{\scE_{\text{out},\omega_n}(\bp_{\rm out})\scE^*_{\text{out},\omega_n}(\bpt_{\rm out})} = \avg{I_{\rm out}} e^{-\omega_n^2 d_2^2 \abs{\bp_{\rm out}-\bpt_{\rm out}}^2 / 32 c^2 L^2} e^{i\omega_n(\abs{\bp_{\rm out}}^2-\abs{\bpt_{\rm out}}^2) / 2 c L},
\end{align}
with $\avg{I_{\rm out}}= P \pi^2 d_1^2 d_2^2 / 64 L^6$. This in turn implies
\begin{align}
	\text{var}(\hat\scP_{\text{out},\dw}(\bp)) = & \frac{\avg{I_{\rm out}}^2}{4\pi^2R^4} \intII{\bp_\text{out}} \intII{\bpt_\text{out}} e^{-(\abs{\bp-\bp_\text{out}}^2 + \abs{\bp-\bpt_\text{out}}^2)/2R^2} \nonumber\\&\times e^{-(\omega_0^2+(\omega_0 + \dw)^2) d_2^2\abs{\bp_\text{out}-\bpt_\text{out}}^2 / 32 c^2 L^2} e^{-i\dw(\abs{\bp_\text{out}}^2-\abs{\bpt_\text{out}}^2) / 2 c L} \\
	= & \frac{\avg{I_{\rm out}}^2 \exp\parens{-\dw^2 R^2 \abs{\bp}^2 / c^2 L^2 Z(R)}}{Z(R)},
	\label{eq:holo-var}
\end{align}
where
\begin{align}
	Z(R)\equiv 1 + \frac{\dw^2 R^4}{c^2 L^2} + \frac{d_2^2 R^2}{8 c^2 L^2} (\omega_0^2+\omega_1^2).
\end{align}
Equation~(\ref{eq:holo-var}) takes its maximum value at $\bp=\bzero$, so we limit our attention to that bound. It is not hard to see that $|\avg{\hat\scP_{\text{out},\dw}(\bp)}|^2 = \avg{I_3}^2$, and so the ratio of the squared mean to the variance is bounded below by
\begin{align}
 	\frac{|\avg{\hat\scP_{\text{out},\dw}(\bp)}|^2}{\text{var}(\hat\scP_{\text{out},\dw}(\bp))} \geq \frac{|\avg{\hat\scP_{\text{out},\dw}(\bp)}|^2}{\text{var}(\hat\scP_{\text{out},\dw}({\mathbf 0}))} = Z(R).
\end{align}
It is clear that the variance vanishes as $R$ grows without bound. However, we would like the variance to be significantly less than the squared mean for a small enough value of $R$ that the $\scP$ field's spatial detail is preserved. Taking reasonable parameter values $\lambda_0=532\nm$, $L=1\m$, $d_2=2\m$, we ambitiously take the difference frequency to be $\dw / 2\pi = 1\,\text{THz}$ so that $\Delta\lambda=300\,{\mu\text{m}}$. Even  taking our Gaussian kernel to have $R=1\,{\mu\text{m}}$, we find $Z(1\,{\mu\text{m}})\approx 141$, implying that the squared mean greatly exceeds the variance, and thus $\hat\scP_{\text{out},\dw}(\bp)\approx\scP_{\text{out},\dw}(\bp)$. Of course, we have traded the need for exceedingly good time resolution in detection for exceedingly good spatial resolution, as sub-micron resolution is required to avoid speckle averaging at the detector arrays.  A possible approach to realizing such arrays is to combine magnifying optics with lock-in cameras~\cite{Foix}, as used by Willomitzer~\emph{et al.}~\cite{smu} in their proofs-of-concept NLoS SWH experiments.

One more point remains to be dealt with, i.e., reconciling Willomitzer~\emph{et al.}'s nonstatistical approach to SWH imaging with the statistics-based $\scP$-field version of SWH imaging that we have just described.  It turns out this will not be hard to do.  Our spatial low-pass filter in Fig.~\ref{fig:SWH}(b) averages out speckle so that $\hat{\scP}_{{\rm out},\dw}(\bp)$ converges to $\scP_{{\rm out},\dw}(\bp)$ \emph{without} unduly compromising spatial resolution. Compare that to the spatial low-pass filtering that Willomitzer~\emph{et al.} report doing---in the discussion below Eq.~(35) of their paper's supplementary material---to suppress ``parasitic interference''.  Their parasitic interference is our speckle, making their (nonstatistical) latent synthetic-wavelength hologram very much like our diffuser-averaged $\hat{\scP}$ field. 

\section*{Funding}
This work was supported by the DARPA REVEAL program under Contract HR0011-16-C-0030.

\section*{Acknowledgments}
The authors acknowledge fruitful interactions with members of the DARPA REVEAL teams from the University of Wisconsin and Southern Methodist University.  They also thank Dr. Ravi Athale for organizing a valuable workshop on phasor-field imaging, and Dr. Jeremy Teichman for sharing an early version of his analysis of speckle effects in $\scP$-field imaging.

\section*{Disclosures}
The authors declare no conflicts of interest.\\


\begin{thebibliography}{99}
\bibitem{footnoteA}Paraxial means that Fresnel diffraction can be used in lieu of the more exact Rayleigh-Sommerfeld diffraction.  A sufficient condition at wavelength $\lambda$ for this to be so is that the longitudinal distance $L$ between the input and output planes and the diameters $d_0$ and $d_L$ of the transverse regions of interest within those planes satisfy $\bar{d}^4/8\lambda L^3 \ll 1$, where $\bar{d} \equiv (d_0+d_L)/2$.  An alternative condition is that the angular spectrum of the input field be confined to a paraxial region about the optical axis.
\bibitem{footnoteB}As shown in Ref.~\cite{speckle}, this will be possible when the image-plane detector's pixels are appreciably smaller than the spatial detail in $\langle I_2(\bp_2,t)\rangle$ and appreciably larger than the speckle size in that plane.
\bibitem{footnote1}The focusing capability of a circular-pupil lens is limited by its diameter $D$.  Our assumption of a Gaussian pupil---here and in all that follows---enables us to obtain closed-form results while still capturing the essential physics of a finite pupil.  In essence, this Gaussian pupil acts as an apodization that suppresses the Fresnel and Fraunhofer diffraction rings which result when a plane wave diffracts through a diameter-$D$ circular pupil while preserving the transition from the near field to the far field.
\bibitem{footnoteC}Physically, $|{\rm Re}[\mathcal{A}(\omega)]| \ll |{\rm Im}[\mathcal{A}(\omega)]|$ is a near-field condition, i.e., it implies that $D$ is much greater than the wavelength-$\lambda_0$ diffraction scale at propagation distance $|1/L_{\rm in}-1/f|^{-1}$ for a diameter-$D$ collimated beam.
\bibitem{footnoteD}The assumption that $|{\rm Re}[\mathcal{B}(\omega)]| \ll |{\rm Im}[\mathcal{B}(\omega)]|$ is a near-field condition similar to the one explained in \cite{footnoteC}.
\bibitem{footnoteE}This input diffuser is not part of the Fig.~1 scenario, i.e., it is not the transmissive analog of any of the walls in three-bounce NLoS imaging.  Instead, it is a user-supplied diffuser whose purpose is to create the initial spatial incoherence that permits $\scP$-field physical-optics projection over a distance $L_{{\rm in}_1}+L_{{\rm in }_2}+L_1$ using an ordinary lens whose focal length satisfies the lens-maker's formula $1/f = 1/L_{{\rm in}_1} + 1/(L_{{\rm in}_2} + L_1)$.
\end{thebibliography}
\end{document}